\documentclass[format=acmsmall, review=false]{acmart}

\usepackage{acm-ec-19}
\usepackage{booktabs} % For formal tables
\usepackage{amsmath}
\usepackage{caption}
\usepackage{subcaption}

\usepackage{color}
\usepackage[textsize=scriptsize]{todonotes}

\definecolor{blue}{RGB}{0, 93, 170}			%Go Big Blue!
\definecolor{darkgreen}{RGB}{0, 102, 0}
\definecolor{myorange}{RGB}{1,0.5,0}

\renewcommand{\citet}[1]{\citeauthor{#1}~\cite{#1}}

\begin{document}
% Title portion. Note the short title for running heads
\title{Heuristics in Multi-Winner Approval Voting}

\author{Jaelle Scheuerman}
\affiliation{Tulane University}
\author{Jason L. Harman}
\affiliation{Louisiana State State University}
\author{Nicholas Mattei}
\affiliation{Tulane University}
\author{K. Brent Venable}
\affiliation{Tulane University and IHMC}

\begin{abstract}
In many real world situations, collective decisions are made using voting. 
Moreover, scenarios such as committee or board elections require voting rules that return multiple winners. 
In multi-winner approval voting (AV), an agent may vote for as many candidates as they wish. Winners are chosen by tallying up the votes and choosing the top-$k$ candidates receiving the most votes. An agent may manipulate the vote to achieve a better outcome by voting in a way that does not reflect their true preferences. In complex and uncertain situations, agents may use heuristics to strategize, instead of incurring the additional effort required to compute the manipulation which most favors them.
%While preliminary results relating to voting heuristics are present in the literature, the conceptualization is often intuitive, not empirical.
In this paper, we examine voting behavior in multi-winner approval voting scenarios with complete information. We show that people generally manipulate their vote to obtain a better outcome, but often do not identify the optimal manipulation. Instead, voters tend to prioritize the candidates with the highest utilities. Using simulations, we demonstrate the effectiveness of these heuristics in situations where agents only have access to partial information.
\end{abstract}

\maketitle

% Renew this after \maketitle if the default list of authors is too long for headers
%\renewcommand{\shortauthors}{W.\ Vickrey et.\ al.}

\section{Introduction}
There are many situations where multiple winners are chosen from a set of candidates by aggregating the preferences of the group using a voting rule. One such rule is approval voting (AV), which is specifically advocated for in multi-winner elections by \citet{Kilg10a}. Approval voting requires that participants vote for zero or more of their favored candidates. The votes are tallied and $k$ winners are chosen based on the total number of votes received by each candidate. Participants may vote with their true preference or vote strategically to obtain a better outcome. Under the basic approval voting rule, an optimal manipulation can be computed in polynomial time when an agent has complete information about the preferences of all the voters \cite{MPRZ08a}. However, under realistic assumptions, voters still may not perform the necessary calculations to identify an optimal outcome. Instead, they may use heuristics, such as voting only for their most preferred candidate or the leader of a poll, to strategize and then cast their vote.

Heuristic decision making has a deep history in psychology and has been used to some degree in the  Computational Social Choice (COMSOC) literature \cite{FLMG19a,ReEn12a}, though this interdisciplinary topic remains for the most part unexplored. In decision science, heuristics can be viewed very differently depending on ones theoretical perspective. One school of thought views heuristics as second best short cuts, when the environment is to complex to use rational strategies \cite{tversky1974judgment}. A second school of thought views heuristics as adaptive strategies that work in the real world, leveraging natural cognitive abilities that exploit the structure of the environment often leading to better outcomes with the use of less information \cite{gigerenzer1999fast}. This second perspective is the one used here. Key to this view of heuristics is that in uncertain environments, decision strategies that ignore some information lead to optimal outcomes.% (i.e. are ecologically rational).\nick{we may need to expand here how this is different fron ECON rational because this is an econ heavy crowd}-It's not really needed, so I just deleted it-Jason

This paper examines the use and effectiveness of heuristics in multi-winner approval voting through an experimental study and simulations. We first define the concepts important to heuristics and voting in Section 2. In Section 3 we examine existing literature relating to approval voting, voting behavior and heuristics. We describe our experiment design and simulation in Section 4 and discuss the results of these in Section 5. Section 6 concludes with a summary of key observations identified and a discussion of future directions.

\section{Preliminaries}
We give a brief overview of the mathematical formalism used to study approval voting and formally define the heuristics of study within this framework.

\subsection{{A}pproval Voting} Following \citet{AGGM+15a} and \citet{Kilg10a} we consider a social choice setting $(N,C)$ where we are given a set $N = \{1, \ldots, n\}$ of voting agents and a disjoint set $C = \{c_1, \ldots, c_m\}$ of candidates.  Each agent $i \in N$ expresses an approval ballot $A_i \subseteq C$ which  gives rise to a set of approval ballots $A = \{A_1, \ldots, A_n\}$.  We study the multi-winner approval voting rule that take as input an instance $(A, C, k)$ and return a subset of candidates $W \subseteq C$ where $|W| = k$ called the winning set.  

The \emph{Approval Voting (AV)} rule finds the set $W \subseteq C$ where $|W| = k$ that maximizes the total weight of approvals (approval score), $AV(W) = \sum_{i\in N} |W \cap A_i|$.  Informally, the winning set under AV is the set of candidates that are approved by the largest number of voters.

In some cases it is necessary to use a tie-breaking rule in addition to a voting rule in order to enforce that the size of $W$ is indeed $k$.  In this study we will fix a tie-breaking rule explicitly, typically we will use the \emph{lexicographic tie-breaking rule} where we assume we are given a fixed ordering over $C$ and we select winners in this order.  Tie-breaking is an important topic in COMSOC and can have significant effects on the complexity of manipulation of various rules even under idealized models \cite{AGM+13a,MNW14a,ObEl11a,OEH11a}.  Investigating the behavior effects of tie-breaking rules, including probabilistic ones, is an interesting direction for future work.

In order to study more closely the behavior of agents and their reactions to incentives, and to align our work with the literature on decision heuristics we assume that each agent $i \in N$ also has a real valued utility function $u_i: C \rightarrow \mathcal{R}$.  We also assume that the utility of agent $i$ for a particular set of winning candidates $W \subset C$ is $u_i(W)=\sum_{c \in W} u_i(c)$ (slightly abusing notation). If $W$ is the subset elected by the voting rule we will refer to $u_i(W)$ as agent's $i$'s \emph{outcome} of the election.

From this model we define a \emph{truthful vote} as one where an agent approves of all candidates for which they have positive utility.  This assumption is made frequently in studies of approval voting \cite{LaSa10a} and is foundational for many of the mathematical models of approval voting \cite{Kilg10a}, i.e., only those candidates with negative or null utility are considered ``unacceptable.'' A \emph{strategic vote} is one in which an agent maximizes their total (expected) utility given a particular decision setting. Note that in some cases, e.g., under the standard approval voting rule when agents only have binary utilities, that a truthful vote is always a strategic vote \cite{MPR08b,MPRZ08a}.  

\subsection{Heuristic Decision Models}
Heuristics are strategies, or adaptive shortcuts, that humans use to make decisions. Originating with the idea of bounded rationality \cite{simon1955behavioral} --- the observation that both the human mind and the environment make application of normative decision models impossible --- heuristics have been studied in depth in a number of fields. Though sometimes seen as second best alternatives when maximization is not possible, other research views heuristics as ideally adapted cognitive abilities that capitalize on the structure of the environment to provide solutions when optimization is not an option, e.g. NP-hard problems, ill defined problems, unfamiliar/time sensitive problems, and also may outperform optimal or maximizing solutions (e.g. heart disease prediction \cite{green1997alters}.)
%Though this literature has a rich taxonomy of heuristics, their building blocks, and the environments where they perform well, 
Key to this idea of heuristics is that they are fast -- composed typically of simple and evolved search, stop, and decision rules -- and they are frugal -- ignoring some, if not much, of casually relevant information. We present two such heuristics inspired from the literature that we believed a priori could be used in multi-winner approval voting, one which ignores vote totals (\textit{take the X best}) and one which ignores preferences (\textit{follow the leader}).

\subsubsection{Take the X best}
The first heuristic considered here is \textit{take the X best}, which is a subset of a truthful vote. It relies on ignoring part of the information available (votes), and uses only  preference magnitude with some cut off \cite{gigerenzer2011heuristic,broder2003take}. Descriptively, this heuristic has two steps: (1) order candidates by preference, (2) select the top-$X$. In this heuristic, X could be calculated a number of ways such as a magnitude cut off or a proportional difference between preferences \cite{brandstatter2006priority}. %Because of the simplified nature of this work, we refrain from positing a specific mechanism and test models with X= 1 and X = 2. 
Because of the preliminary nature of the current work, we do not assume a specific stop rule, opting instead to test all possible sizes of X. More formally, When an agent votes with the \textit{take the X best} heuristic, they vote for a subset $P$ of preferred candidates of size $X$, ordered by their utility. Formally, $P = p_1 > ... > p_x$ where $u_1(p_1) > ... > u_X(p_X)$. $|P| = X$ and must be at least 1 and fewer than the total number of candidates with positive utility. For example, if there are four candidates with positive utility, \textit{take the X best} could refer to voting for the candidate with the highest utility, up to the top three candidates.

One may consider our definition of a truthful vote, i.e., that an agent votes for all candidates for which he has a positive utility, too restrictive and that most agents would most often use some form of take the X best, especially when there are large discrepancies between the utility of various candidates.  While this is a valid criticism and we hope to tease out this relationship in future work; we again observe that many of the axiomatic studies of approval voting (and its variants) require the assumption that agents approve all agents with positive utility \cite{Kilg10a,AGGM+15a}.  Hence, rather than this being a weakness of our work, we see this as a good jumping off point in investigating what happens when mathematical assumptions enter the messy world of human decision making.

\subsubsection{Follow the leader}
Follow the leader is a heuristic that ignores preference information and bases choice only on social cues. It has been shown to be an effective heuristic in both spatial movement and cue order learning \cite{seitz2016cognitive}. When the \textit{follow the leader} heuristic is used, agents include one or more candidates with the most votes in the voting profile, whether or not 
such candidates have positive utilities.

\section{Related Work}

Approval voting is a set of methods for aggregating group preferences that are particularly popular among economists, computer scientists, psychologists, and beyond \cite{LaSa10a,BrFi07c}.  There are even multiple political action committees (PACs) in the United States, e.g., The Center for Election Science\footnote{\url{https://www.electionscience.org/}}, that are committed to seeing the United States change voting procedures from plurality to approval voting.  Perhaps one reason for this popularity is the idea that participants are allowed to express preference over a set of candidates and not just a single one.  In France, a large study was run parallel to the 2002 election showing that many voters would have preferred approval ballots to traditional plurality ballots \cite{LaVa08a}.

%Academic studies of approval voting (AV) include numerous studies about 
The complexity of manipulation for various types of approval voting (AV) has received considerable attention in the Computational Social Choice (COMSOC) literature \cite{BCELP16a}.  COMSOC is an area of research that investigates computational issues surrounding the aggregation of individual preferences.  To start, assuming that agents act rationally and have full information about the votes of other agents, when agents have \emph{Boolean utilities}, i.e., when all agents either have utility 1 or 0 for candidates they approve or disapprove of, respectively, AV is strategy-proof.  When agents have general utilities, finding a vote that maximizes the agent's utilities can be computed in polynomial time \cite{MPR08b,MPRZ08a}.  For variants of AV including Proportional Approval Voting, Satisfaction Approval Voting, and the Repeated Approval Voting, the complexity of finding utility maximizing votes, even if they are manipulations, ranges in complexity from easy to coNP-complete \cite{AGGM+15a}.

%(see \citet{AGGM+15a} for a full discussion of the computational issues of these variants of approval voting).

Many theoretical works in COMSOC make worst case computational assumptions: manipulators have complete information, all votes are know, etc.  However, there are several efforts within the COMSOC community to expand these worst case assumptions and strategic issues are investigated in the presence of uncertain information or when agents are not perfectly rational.  In \citet{ReEn12a}, agents are given access to poll information and agent behaviors are modeled as being $k$-pragmatist, i.e., they only look at the top $k$ candidates when deciding whether or not to make strategic decision.  In \citet{MLR14a}, agents are modeled as behaving in \emph{locally dominant} ways, i.e., they take into account only a small number of possible outcomes when deciding whether or not to act strategically in a particular voting setting.  A survey of other recent work in COMSOC specifically on issues surrounding strategic voting is given by \citet{Meir18a}.
 
The main tools are theory and, more recently, simulation to investigate the behavior of agents. There is a growing effort to use real-world data and simulation to test various decision making models, e.g., \cite{MaWa17,MaWa13a}. Within the economics and psychology literature there have been several studies of approval voting and the behavior of voters, however, these studies typically use data from past elections and use re-sampling or other methodologies to look at possible effects of strategic voting \cite{FMR93a,MFG12a,Matt11a,RHT07a}.  Perhaps the most interesting and relevant to our work is the studies of \citet{RHT07a} which focus on elections of various professional societies where approval balloting was used and the work of \citet{ZMD15a} where many approval voting settings were obtained from Doodle, an online polling platform.  In \citet{RHT07a} election data is used along with proposed heuristics for individual choice behavior, the conclusion is that many voters use an \emph{plurality heuristic} when voting in AV elections, though they may not be using this heuristic to strategic ends. In \citet{ZMD15a}, analysis of the Doodle data revealed that when the participants can see the votes of others they are more likely to approve more meeting times; and that there is a statistically significant difference in the behavior of agents between closed and open polls.

In this paper we begin to move beyond simply simulation and re-purposed data to studying the preferences and decision making behavior of actual decision makers in realistic settings \cite{ACGM+15a}.  This goal, of collecting focused data, is perhaps most well known from the field of experimental economics \cite{KaRo95a} and behavioral game theory \cite{Came11a}.  Essentially, we can use payments to individuals to study how they would act in real-world settings.  There are a number of best practices for providing appropriate incentives to individuals or over crowd-working platforms to study decision making \cite{MaSu12a,HSSW15a}.

Three recent papers both address strategic voting under the plurality rule where agents are making decisions in uncertain environments.  First, \cite{TySc16a} study the voting behavior of agents under the plurality rule with three options.  They find that the amount of information available to the voters affects the decision on whether or not to vote strategically and that in many cases the strategic decisions do not affect the outcome of the plurality vote.

Second, in \citet{TMG15a} an online system is presented where participants vote for cash payments in a number of settings using the plurality rule under uncertainty. Two specific scenarios are studied: one where a user votes after being given access to a large pre-election poll and the second where agents vote simultaneously and can update their votes.  They find that most participants do not engage in strategic voting unless there is a clear way to benefit.  %Finding that typically when agents are presented with uncertain poll information, when agents are strategic, ~70\% of voters will vote for their second most preferred candidate when their most preferred candidate has no or a small chance of winning, when their most preferred candidate is ranked second a large proportion vote for the less preferred candidate. 
In the iterative setting most voters were lazy and if they did vote strategically, they would do a one step look ahead or perform a best response myopically.

Finally, in \citet{FLMG19a} a comprehensive study using both past datasets and newly collected ones examines the actual behavior of agents in multiple settings with uncertainty versus behavior that is predicted by a number of behavioral and heuristic models.  The paper includes a proposal for a novel model of user voting behavior in these uncertain settings called \emph{attainable utility}, where agents consider how much utility they would gain versus the likelihood of particular agents winning given an uncertain poll.  They conclude that the attainable utility model is able to explain the behavior seen in the experimental studies better than existing models and even perform near the level of state of the art machine learning algorithms in modeling users' actual behavior.

There are many other closely related studies of strategic voting and preference reporting in cooperative and noncooperative settings.  Recently, \citet{MWPS15a} developed a large system to look at how users report preferences to an allocation mechanism and studied the strategic behavior of those users; finding that most users were myopic and that searching a small space of \emph{local strategies} accounted for most strategic voting.  There is also a series of papers looking at predicting and modeling user behavior in simple games, including voting games \cite{WrLe14a,HaWrLe16,WrLe12}.  In each of these papers large corpora of humans playing simple games are collected and then models are proposed to fit the observed behavior.  Generally speaking, most participants are only strategic in a limited sense and do not look ahead as far into the future as one would expect based on formal game theoretic models \cite{MSZ13a}.Other recent work with human decision making in cooperative games settings includes weighted voting games \cite{MBZ17a} and coalition allocation games \cite{DLL19a}.  

\section{Experiment design}

In this section we describe the design of two experiments aimed at studying the role of heuristics in voting, respectively, when there is opportunity for manipulation and when there is uncertainty about votes.

\subsection{Study 1: Heuristics vs Manipulation}

Four different scenarios were designed to highlight how the heuristics may be used in multi-winner approval voting. Each scenario includes a set of five candidates, the agent's utility function, and the total votes for each candidate so far; see the top table of Figure \ref{fig:scenario1-des} for an example. The agent is assumed to be the last voter and can vote for between 0-5 candidates. The winning set includes $k$ winners with the highest number of votes. 
%A second between-subjects manipulation was the number of winners in an election.
The four voting scenarios were presented to participants as either elections where two candidates would win or three candidates would win. To ensure that the winning set contains at most $k$ candidates, ties are broken lexicographically, with $A > B > C > D > E$. After data collection, voting profiles were categorized as strategic (i.e., optimal manipulation), truthful,  heuristic (i.e., \textit{take the X best} and  \textit{follow the leader}), or other.  

\subsubsection{Scenario 1} In the first scenario, shown in the top-table of Figure \ref{fig:scenario1-des}, the agent's utility function includes a positive utility for candidates $A$, $B$ and $E$.  As summarized in the bottom table of Figure \ref{fig:scenario1-des}, voting truthfully for all three candidates results in a sub-optimal outcome, with a total utility of 0.15. In the two-winner condition, using the \textit{take the X best} heuristic with $X=1$, i.e.,  voting only for the highest utility candidate, achieves the optimal outcome with a utility of 0.25. Voting for the top two candidates results in an outcome worse than truthful. In the three-winner condition, using \textit{take the X best} with $X=2$, i.e., voting for the top two candidates, results in the  optimal outcome with utility of 0.35.

\begin{figure}[h]
     \centering 
\begin{tabular}{|c c c c c c|}
     \hline
        \textbf{Candidates:} & \textbf{A} & \textbf{B} & \textbf{C} & \textbf{D} & \textbf{E}  \\
     \hline
        \textbf{Utilities:}  & 0.05 & 0.10 & 0 & 0 & 0.25  \\ 
     \hline
        \textbf{Vote Totals:} & 3 & 3 & 4 & 3 & 3 \\
    \hline
     \end{tabular}
     \vspace{0.5cm}

\begin{tabular}{ |c |c |c | c| }
\hline
  \textbf{Strategy} & \textbf{Voting Profile} & \textbf{Two winners} & \textbf{Three winners} \\ 
 \hline
 Truthful & [A,B,E] & 0.15 & 0.15 \\ 
 Take the X best (X=1) & [E] & \textbf{0.25} & 0.30 \\  
 Take the X best (X=2) & [B,E] & 0.10 & \textbf{0.35} \\
 \hline
\end{tabular}

    \caption{\label{fig:scenario1-des} Scenario 1: Design. Subject's view (top table) and resulting utilities for Scenario 1 strategies (bottom table). Optimal strategies are in bold.}
\end{figure}

\subsubsection{Scenario 2} The second scenario, shown in Figure \ref{fig:scenario2-des}, is similar to the first, except that the leading candidate provides a very small utility if elected. This scenario examines whether people choose to vote truthfully, \textit{follow the leader} when the leader has some small utility, or manipulate their vote to obtain a better outcome. The utilities that result from each of these strategies are summarized in the table at the bottom of  Figure \ref{fig:scenario2-des}. If subjects choose to \textit{follow the leader} and vote only for C, the resulting utility is identical to that of voting truthfully. Voting for the highest utility candidate results in the optimal outcome for the two-winner condition and voting for the top two candidates results in the optimal outcome for the three-winner condition. We note here that voting for the three highest utility candidates could also be interpreted as truthful voting that discounts very small utilities.

\begin{figure}[h]
     \centering 
 \begin{tabular}{|c c c c c c|}
     \hline
        \textbf{Candidates:} & \textbf{A} & \textbf{B} & \textbf{C} & \textbf{D} & \textbf{E}  \\
     \hline
        \textbf{Utilities:}  & 0.05 & 0.10 & 0.01 & 0 & 0.25  \\ 
     \hline
        \textbf{Vote Totals:} & 3 & 3 & 4 & 3 & 3 \\
     \hline
     \end{tabular}
     
     \vspace{0.5cm}
     
%\begin{table}[h]
\begin{tabular}{ |c |c |c | c| }
\hline
  \textbf{Strategy} & \textbf{Voting Profile} & \textbf{Two winners} & \textbf{Three winners} \\ 
  \hline
 Truthful & [A,B,E] & 0.06 & 0.16 \\ 
 Take the X best (X=1) & [E] & \textbf{0.26}  & 0.31 \\  
 Take the X best (X=2) & [B,E] & 0.11 & \textbf{0.36} \\
 Take the X best (X=3) & [A,B,E] & 0.15 & 0.16 \\
 Follow the Leader & [C] & 0.06 & 0.16 \\
 \hline
\end{tabular}
%\end{table}

 \caption{\label{fig:scenario2-des} Scenario 2: Design. Subject's view (top table) and resulting utilities for Scenario 2 strategies (bottom table). Optimal strategies are in bold.}
    
\end{figure}

\subsubsection{Scenario 3} In the third scenario, shown in Figure \ref{fig:scenario3-des}, the voting profiles are designed so that two candidates with no utility have a large lead over the remaining candidates. In the two-winner condition, this results in a no-win scenario. We expect that more people will choose to vote truthfully or use the \textit{follow the leader} heuristic. In the three-winner condition, an optimal outcome with utility of 0.25 can be obtained using the \textit{take the X best heuristic} with $X=1$. Outcomes with optimal utility can also be obtained by voting for the candidate with the highest utility together with one or both of the leading candidates. These strategies are summarized in the table at the bottom of Figure \ref{fig:scenario3-des}.

\begin{figure}[h]
     \centering 
     
\begin{tabular}{|c c c c c c|}
     \hline
        \textbf{Candidates:} & \textbf{A} & \textbf{B} & \textbf{C} & \textbf{D} & \textbf{E}  \\
     \hline
        \textbf{Utilities:}  & 0.05 & 0.10 & 0 & 0 & 0.25  \\ 
     \hline
        \textbf{Vote Totals:} & 1 & 1 & 4 & 4 & 1 \\
     \hline
     \end{tabular} 
\vspace{0.5cm}

%\begin{table}[h]
\begin{tabular}{ |c |c |c | c| }
\hline
  \textbf{Strategy} & \textbf{Voting Profile} & \textbf{Two winners} & \textbf{Three winners} \\ 
  \hline
 Truthful & [A,B,E] & 0 & 0.05 \\ 
 Take the X best (X=1) & [E] & 0  & \textbf{0.25} \\  
 Take the X best (X=2) & [B,E] & 0 & 0.10 \\
 Follow the Leader  & [C] & 0 & 0 \\
 Follow the Leader + Take the best & [C,E] or [C,D,E] or [D,E] & 0 & \textbf{0.25} \\
 \hline
\end{tabular}
%\caption{\label{tab:scenario3} Resulting utilities for Scenario 3 strategies. Optimal strategies are in bold.}
%\end{table}

\caption{Scenario 3: Design. Subject's view (top table) and resulting utilities for Scenario 3 strategies (bottom table). Optimal strategies are in bold.}
    \label{fig:scenario3-des}
\end{figure}

\subsubsection{Scenario 4} The final scenario, shown in Figure \ref{fig:scenario4-des},  includes a candidate with a large negative utility. In this situation, we expected that subjects would prioritize ensuring that D does not win over some other heuristic. In the two-winner condition, a sub-optimal outcome that did not include D, could be achieved by voting only for the highest utility candidate, whereas the optimal outcome could be achieved by voting for the top two candidates. In the three-winner scenario, being truthful results in the optimal outcome and using the \textit{take the X best} heuristic leads to an outcome where the disliked candidate wins. Utilities are summarized in the bottom Table of Figure \ref{fig:scenario4-des}.

\begin{figure}[h]
     \centering 
\begin{tabular}{|c c c c c c|}
     \hline
        \textbf{Candidates:} & \textbf{A} & \textbf{B} & \textbf{C} & \textbf{D} & \textbf{E}  \\
     \hline
        \textbf{Utilities:}  & 0.05 & 0.10 & 0 & -1 & 0.25  \\ 
     \hline
        \textbf{Vote Totals:} & 3 & 3 & 4 & 4 & 3 \\
     \hline
     \end{tabular} 
     
     \vspace{0.5cm}

%\begin{table}[h]
\begin{tabular}{ |c |c |c | c| }
\hline
  \textbf{Strategy} & \textbf{Voting Profile} & \textbf{Two winners} & \textbf{Three winners} \\ 
  \hline
 Truthful & [A,B,E] & 0.15 & \textbf{0.40} \\ 
 Take the X best (X=1) & [E] & 0.25  & -0.75 \\  
 Take the X best (X=2) & [B,E] & \textbf{0.35} & 0.35 \\
 \hline
\end{tabular}
%\caption{\label{tab:scenario4} Resulting utilities for Scenario 4 strategies. Optimal strategies are in bold.}

\caption{\label{fig:scenario4-des} Scenario 4: Design. Subject's view (top table) and resulting utilities for Scenario 4 strategies (bottom table). Optimal strategies are in bold.}
    
\end{figure}
%\end{table}

\subsubsection{Experiment Implementation}
\paragraph{Participants.} 129 undergraduate students completed the study either voluntarily or for course credit at two different universities. 
\paragraph{Procedure.} The study was a 1 within (scenario) X 2 between (2 or 3-winners) repeated design, with 66 subjects in the 2-winner condition and 63 in the 3-winner condition. After giving informed consent, participants read instructions which explained approval voting and the tie breaking mechanism with examples. They then advanced through the four scenarios, selecting which candidates they voted for before moving on to the next election. Each election displayed the candidates, the number of votes for each candidate before the participants final vote, and the utilities for each candidate, as shown in the top tables of Figures \ref{fig:scenario1-des}, \ref{fig:scenario2-des}, \ref{fig:scenario3-des} and \ref{fig:scenario4-des}. All outcomes were hypothetical and participants were told that the utilities were monetary representations of how much they preferred a candidate.  
The first condition involved scenarios with two winners, while the second condition examined scenarios with three winners. When voting, subjects could choose to vote truthfully (for all candidates with positive utility), or manipulate their vote to achieve a higher utility. For this experiment, it was expected that most people would try to vote strategically, but that they would not perform all of the necessary computations to identify the optimal outcome. Instead, we expected that people would use heuristics to identify voting profiles that increase their utility over being truthful.

\subsection{Study 2: Heuristics in the Presence of Uncertainty}

The scenarios and experiment described above were designed to simulate voting conditions with complete information. 
One of the goals of the science of heuristics is to identify under what circumstances (e.g. structure of uncertainty) are heuristic strategies as good or better than optimal (or maximizing) strategies. To begin to explore this question in approval voting we ran simulations  for the  scenarios above while introducing uncertainty.  We could have structured uncertainty in at least three different ways, each of which may be more or less ecologically valid (and hence generalizable). The first is uncertainty in information, i.e. the vote totals provided, as is the case with pre-election polls we could introduce margins of error in the vote totals. A second form of uncertainty is in the selection mechanism. We used a lexicographic tie breaking rule, where if there were ties they were broken alphabetically. An alternative would have been random selection, where ties are broken by coin flip. A third form of uncertainty could be introduced via uncertainty of the future. In this case, we could create a scenario where the participant is not the last person to cast a vote, but next to last, or two back etc. In the later two examples, the ultimate winners may not be knowable a priori, but probabilities are easily estimated and optimal solutions can be calculated. There are no doubt more ways to conceive of uncertainty in approval voting but these three are apparent. To begin to explore the effectiveness of different strategies under uncertainty we focused on the tie breaking rule and subsequent unknown votes.
%in a series of simulations for study 2.  
%Understanding how effective different heuristics are in situations with uncertainty represents an important next step. 
A simulation was developed to identify the optimal strategies when uncertainty is present. Uncertainty was added to Scenarios 1-4 in the two forms described above. 
%First, the ties were broken randomly, instead of lexicographically. A second source of uncertainty was added by extending the scenarios so that there are one, two or three voters remaining after the agent's vote is cast. 
We calculate the probability of every possible set of aggregated votes and the expected utilities for each voting profile. Using these expected utilities we are able to demonstrate the effectiveness of different heuristic strategies under different types and levels of uncertainty (see Section 5.2).

\section{Results}
\subsection{Study 1}
Results from Study 1 showed three distinct patterns of voting profiles; truthful, optimal, and heuristic. Across both conditions only 12.40\% of participants voted for all of the the candidates with positive utilities, indicating that a majority of participants did attempt to manipulate the outcome of each election. 24.22\% of the voting profiles across all elections in both conditions were optimal, while 55.82\% of the voting profiles fit into one of the pre-identified heuristics.
%(take the X best or follow the leader).  

\subsubsection{Scenario 1.} Figure 5 shows the stimuli for the first election along with the profile frequencies for the two and three winner conditions. Data in the two winner condition was uni-modal, with the largest proportion of participants voting for Candidate E (n = 16, $24.24\%$, earned utility = $0.25$), which was the optimal and also fits the \textit{take the X best} heuristic. Truthful voting has the next highest frequency (n = 10, $15.15\%$, earned utility = $0.15$). 
Likewise, in the three winner condition a majority of participants voted for B and E (optimal, take the X best: n = $15$, $23.81\%$, earned utility = 0.35). This was followed by truthful voters having the second highest frequency (n = $13$, $20.63\%$, earned utility = $0.15$).

\begin{figure}[h]

     \centering 
     
     \begin{tabular}{|c c c c c c|}
     \hline
        \textbf{Candidates:} & \textbf{A} & \textbf{B} & \textbf{C} & \textbf{D} & \textbf{E}  \\
     \hline
        \textbf{Utilities:}  & 0.05 & 0.10 & 0 & 0 & 0.25  \\ 
     \hline
        \textbf{Vote Totals:} & 3 & 3 & 4 & 3 & 3 \\
    \hline
     \end{tabular}
    
     \begin{subfigure}[b]{0.8\textwidth}
        \centering
        \includegraphics[width=\textwidth]{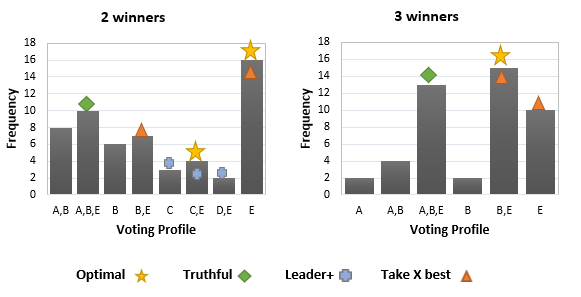} 
    \end{subfigure}
    
     \caption{Scenario 1. Stimuli and observed profile frequencies are displayed with icons indicating which strategy each profile is categorized as.}
    \label{fig:scenario1}
\end{figure}

Results for Scenario 1 show that a majority of participants did not vote truthfully. Because the optimal vote in each condition overlaps with \textit{take the X best}, differences between the two cannot be disentangled for this scenario.

\subsubsection{Scenario 2.} Figure 6 shows the stimuli for the second election along with the profile frequencies for the two and three winner conditions. The second scenario was identical to the first with the exception of a very small utility added to the leading candidate. Data in the two winner condition was similar to that of scenario 1 where the modal response was voting for Candidate E (n = $17$, $25.76\%$, earned utility = $0.26$) which was the optimal and also fits the \textit{take the X best} heuristic. The second highest frequency was voting for B and E (take the X best: n = $9$, $13.64\%$, earned utility = $0.11$) followed by A and B (n = $8$, $12.12\%$, earned utility = $0.15$). 
In the three winner condition a majority of participants voted for B and E (optimal, take the X best: n = $14$, $22.22\%$, earned utility = $0.36$) with A, B, and E (take the X best, truthful - low utility) having the second highest frequency (n = $12$, $19.05\%$, earned utility = $0.16$).

\begin{figure}[h]

     \centering 
     
     \begin{tabular}{|c c c c c c|}
     \hline
        \textbf{Candidates:} & \textbf{A} & \textbf{B} & \textbf{C} & \textbf{D} & \textbf{E}  \\
     \hline
        \textbf{Utilities:}  & 0.05 & 0.10 & 0.01 & 0 & 0.25  \\ 
     \hline
        \textbf{Vote Totals:} & 3 & 3 & 4 & 3 & 3 \\
     \hline
     \end{tabular} 
     
     \begin{subfigure}[b]{0.8\textwidth}
        \centering
        \includegraphics[width=\textwidth]{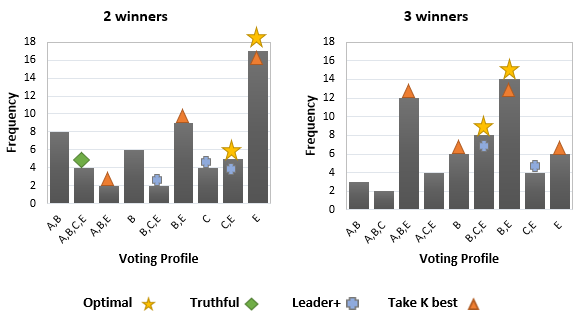} 
    \end{subfigure}
    
     \caption{Scenario 2. Stimuli and observed profile frequencies are displayed with icons indicating which strategy each profile is categorized as.}
    \label{fig:scenario2}
\end{figure}

In Scenario 2, results were similar to the first scenario with modal responses being voting profiles that were optimal but also fit \textit{take the X best}. 4 participants voted truthfully with the low utility candidate (all in the two winner condition) while 14 (12 in the three winner condition) voted truthfully for non-trivial utilities (discounting C).  

\subsubsection{Scenario 3.} Figure 7 shows the stimuli for the third election along with the profile frequencies for the two and three winner conditions. The third scenario was designed as a no-win setting in the two winner condition, with the two non-preferred candidates having insurmountable leads. Voting profiles for the two winner condition are relatively uniform with no clearly dominant profile. Of the profiles with more than 5 observations are E (take the X best), B and E (take the X best), C and D (follow the leader), and A, B, and E (truthful). In the three winner condition, where one preferred candidate can win, the dominating profile (n = $19$, $30.16\%$) was voting for E (optimal, take the X best, earned utility = $0.25$) With the next highest frequency being truthful (n = $10$, $15.87\%$, earned utility = $0.05$).

\begin{figure}[h]

     \centering 
     
     \begin{tabular}{|c c c c c c|}
     \hline
        \textbf{Candidates:} & \textbf{A} & \textbf{B} & \textbf{C} & \textbf{D} & \textbf{E}  \\
     \hline
        \textbf{Utilities:}  & 0.05 & 0.10 & 0 & 0 & 0.25  \\ 
     \hline
        \textbf{Vote Totals:} & 1 & 1 & 4 & 4 & 1 \\
     \hline
     \end{tabular} 
     
     \begin{subfigure}[b]{0.8\textwidth}
        \centering
        \includegraphics[width=\textwidth]{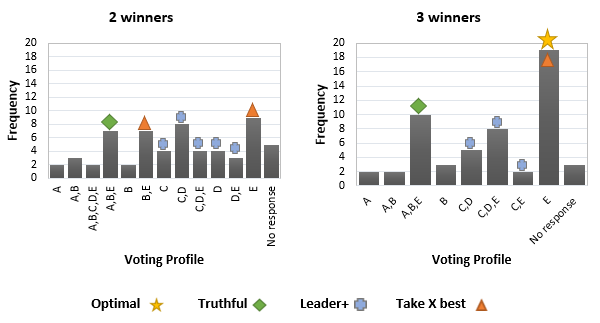} 
    \end{subfigure}
    
     \caption{Scenario 3. Stimuli and observed profile frequencies are displayed with icons indicating which strategy each profile is categorized as.}
    \label{fig:scenario3}
\end{figure}

In the no-win condition of Scenario 3 results were, as expected, widely distributed. Five participants voted for no candidate, while each frequency higher than that fit into one of the pre-identified heuristics or were truthful. In the three winner condition, profile patterns were between optimal, \textit{take the X best} and truthful. Across both conditions a non-trivial number of participants (n = $13$, $10.01\%$) voted only for the leaders, which had zero utility. 

\subsubsection{Scenario 4.} Figure 8 shows the stimuli for the fourth scenario along with the profile frequencies for the two and three winner conditions. The fourth scenario introduced negative utility for Candidate D. Data in the two winner condition showed a clear majority of participants voting for E, which is consistent only with \textit{take the X best} (n = $18$, $27.27\%$, earned utility = $0.25$) with the second highest frequency being for B and E (optimal, take the X best: n = $10$, $15.15\%$, earned utility = $0.35$). In the three winner condition, truthful voting was also optimal and had the highest frequency (n = $17$, $26.98\%$, earned utility = $0.40$).

\begin{figure}[h]

     \centering 
     
     \begin{tabular}{|c c c c c c|}
     \hline
        \textbf{Candidates:} & \textbf{A} & \textbf{B} & \textbf{C} & \textbf{D} & \textbf{E}  \\
     \hline
        \textbf{Utilities:}  & 0.05 & 0.10 & 0 & -1 & 0.25  \\ 
     \hline
        \textbf{Vote Totals:} & 3 & 3 & 4 & 4 & 4 \\
     \hline
     \end{tabular} 
     \begin{subfigure}[b]{0.8\textwidth}
        \centering
        \includegraphics[width=\textwidth]{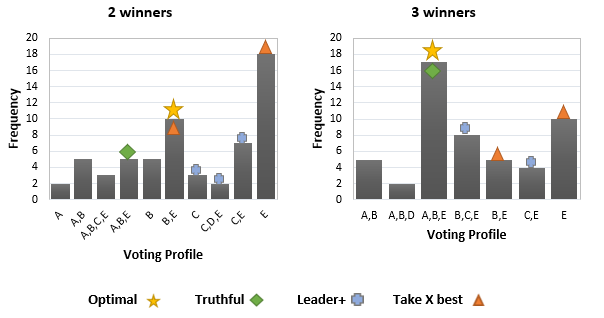} 
    \end{subfigure}
    
     \caption{Scenario 4. Stimuli and observed profile frequencies are displayed with icons indicating which strategy each profile is categorized as.}
    \label{fig:scenario4}
\end{figure}

The introduction of possible losses in Scenario 4 provided some of the more interesting results across all scenarios. The two winner condition is the first where a \textit{take the X best} profile (X=1) is clearly distinct from and preferred to the optimal solution, which if chosen would have increased earnings by $40\%$. Frequencies were nearly reversed in the three winner condition where the optimal is also truthful and \textit{take the X best} would result in overall negative earnings.

Overall results from looking at profile frequencies across conditions leads to three preliminary observations; 1) participants do not uniformly vote truthfully, 2) optimal manipulation is not a dominant strategy, and 3) heuristics such as \textit{take the X best} do capture a significant proportion of voting profiles. Next we explore differences between two winner and three winner elections and examine the data at the individual level, across all four scenarios to see if there are individual consistencies in voting profiles (i.e. consistent maximizers or truthful voters).

\subsubsection{Additional Analysis}
Changing a scenario from two to three winners changes the optimal solution and hence direct comparisons are not straightforward. Overall, participants voted for more candidates in a profile (M = 2.17 candidates) in the three winner condition than the two winner condition (M = 1.75 candidates). In terms of perceived complexity, across elections where an optimal solution exists 28.97\% of profiles were optimal in the three winner condition while 26.26\% were optimal in the two winner condition. Truthful votes were distributed 15.87\% for three winner and 9.09\% in two winner elections, and profiles consistent with \textit{take the X best} were split 38.49\% and 36.74\% for the three and two winner elections respectively. 

Our experimental design allows us to look at within subjects patterns of voting to perform a type of policy capturing where we can identify participants who tend to follow one voting pattern consistently or not. We do find evidence for clusters of participants who have clear voting strategies. For truthful voters, voting for all candidates that have positive utility, no participants voted in this manner, but 5 voted truthfully in at least 3 of 4 elections. Of those 5, two voted for all candidates with non-trivial utilities in every election. In terms of maximizers, 5 participants voted for the optimal solution in every election that had an optimal solution, while 14 voted optimally in at least 3(2) of the 4(3) qualifying elections.  \textit{Take the X best} was used consistently by 9 participants and an additional 12 participants had a voting profile consistent with \textit{take the X best} on at least 3 of the 4 elections. 

Thus far, we have shown that most individuals in this context do not vote truthfully for the options they prefer (i.e. have positive utility) but instead vote in another manner. We have identified optimal and heuristic (\textit{take the X best})  as two alternative strategies to voting truthfully. What are the practical implications for these different strategies?  We can view the outcome of each strategy by calculating the average earnings if a participant were to rely on one strategy across elections. Doing this, the optimal strategy obviously had the highest average earning which are 1.11 on average if a participant voted for the optimal profile in each election. If however, a participant used a single heuristic (voting the same across two and three winner elections), \textit{take the X best} with (X=2) would have average earnings of 0.86, followed by truthful voting with 0.59. \textit{Take the X best} with X=1 had the lowest average earnings in these elections and would earn 0.29 averaging across two and three winner conditions. In this instance with a clear optimal solution, the heuristic could be considered second best. However, the world is rarely set up where an optimal solution is easily identifiable with no uncertainty, and optimal strategies often have to rely on expected utility type calculations, which can be cumbersome if not impossible. Heuristics on the other hand are adaptive and, beyond minimizing effort, actually have the potential to yield results better than optimal strategies in more naturalistic environments with realistic uncertainty \cite{gigerenzer1996reasoning}.

\subsection{Study 2}
Simulations of voting behavior under uncertainty were completed for both two-winner and three-winner conditions in each of the four scenarios described in Section 4.1. The simulation calculated the expected utility of all possible voting profiles assuming that ties were broken randomly and that 0, 1, 2 and 3 voters would remain after the vote was cast. The following figures plot the expected utility across different levels of uncertainty for four voting profiles along with the maximum utility possible (Max). The profiles included in the simulations are; \textit{take the X best} heuristic with X=1 and X=2, truthful voting, and static maximization (Max*) which is a maximizing profile when there are no future votes. This final profile (Max*) represents a maximization strategy when some uncertainties are unknown as is the case in many dynamic environments. Examining the resulting expected utilities shows a pattern where using a \textit{take the X best} heuristic provides the maximum expected utility in 5 of the 8 simulations. Maximizing assuming certainty (Max*) provides the maximum utility in 3 of the 8 simulations and is outperformed by \textit{take the X best} in 4 of 8 simulations, while truthful voting is optimal in 1 of 8 simulations.  

\subsubsection{Simulations for Scenario 1.} Figure \ref{fig:sim1} shows the expected utilities for scenario 1 for different strategies, including truthful voting, static maximizing, and \textit{take the X best}. The maximum expected utility  with any strategy is also identified. In  both the two-winner and three-winner conditions, voting truthfully results in a lower expected utility than using a \textit{take the X best} heuristic. In the two-winner condition, the optimal strategy can be obtained by voting only for the top candidate, when the voter is the last to vote under the random tie-breaking rule. As more uncertainty is added to the scenario, it becomes preferable to take the top two candidates. In the three-winner condition, it is always preferable to vote for the top two candidates. The maximizing profile when no voters are remaining quickly becomes disadvantageous as more uncertainty is added, becoming the worst profile when two or more voters remain.

\subsubsection{Simulations for Scenario 2.} 
\begin{figure}[h]
    \centering 
        \begin{subfigure}[b]{0.8\textwidth}
            \centering
            \includegraphics[width=\textwidth]{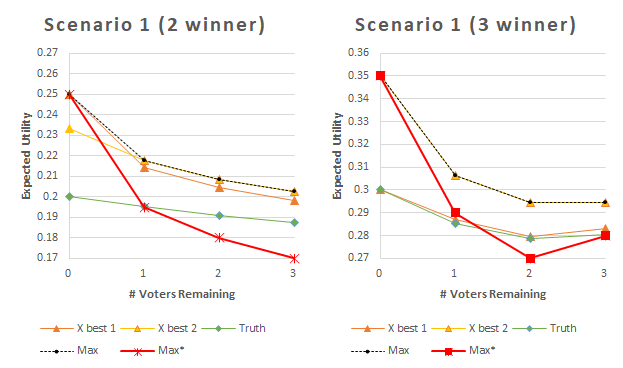}
        \end{subfigure}
        
        \begin{subfigure}[b]{0.8\textwidth}
            \centering
            \includegraphics[width=\textwidth]{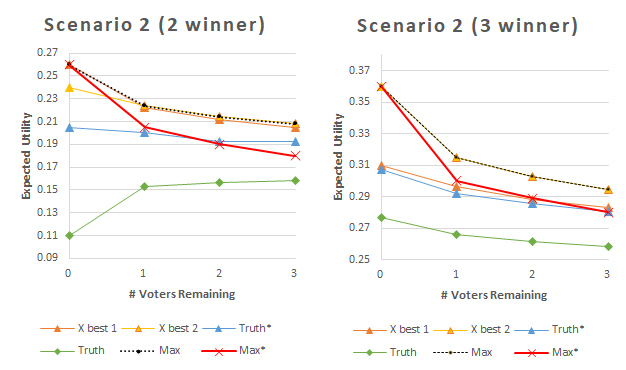}
        \end{subfigure}
        \caption{Simulation results for scenarios 1-2.}
        \label{fig:sim1}
\end{figure}

The expected utilities for scenario 2 are shown in Figure \ref{fig:sim1}. Like Scenario 1, voting truthfully results in a lower expected utility than using a \textit{take the X best} heuristic. Voting for all but the candidate with a trivial utility (truth*) results in a better outcome than being truthful, but voting for the two most preferred candidates results in the maximum expected utility in both conditions. The maximizing profile at 0 voters remaining again drops below \textit{take the X best} as uncertainty is added but remains preferable to truthful voting and is equivalent to truth* in both conditions and \textit{take the X best} with X=1 in the 3 winner condition. 

\subsubsection{Simulations for Scenario 3.} The simulations for scenario 3, shown in Figure 10, provide some different results between the two-winner and three-winner condition, which are shown in Figure \ref{fig:sim1}. The scenario was designed so that the two-winner condition resulted in a no-win situation when there are 0 or 1 voters remaining. If two or more voters are remaining, it is possible for this scenario to generate a winner in the preferred set. In the three-winner scenario, using the \textit{take the X best} strategy will elect the voter's favored candidate in addition to the leading candidates with no utility. The two-winner condition creates an environment where using the \textit{take the X best} heuristic can be harmful and voting truthfully results in the maximum possible expected utility. When voting for the top two candidates, the utilities are almost as good as voting truthfully, but the expected utility goes down when voting for just the top candidate. The three-winner condition results in the same pattern seen in scenarios 1 and 2, where voting for the top-X candidates results in a higher expected utility over being truthful. In this situation, voting for the single highest utility candidate is the optimal choice. There is a static maximizing profile in the two-winner condition and in the three-winner condition which is identical to \textit{take the X best} with X=1.

\begin{figure}[h]
    \includegraphics[width=0.8\textwidth]{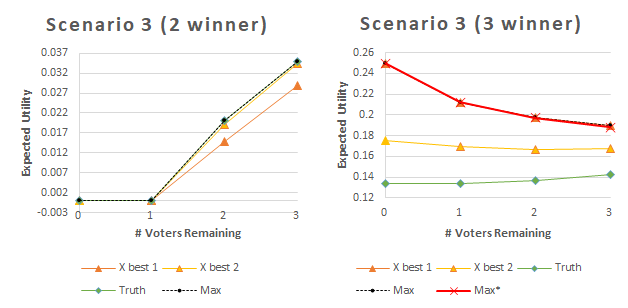}
        \caption{Simulation results for scenario 3.}
\end{figure}

\subsubsection{Simulations for Scenario 4.} 
\begin{figure}[h]
    \centering 
        \includegraphics[width=0.8\textwidth]{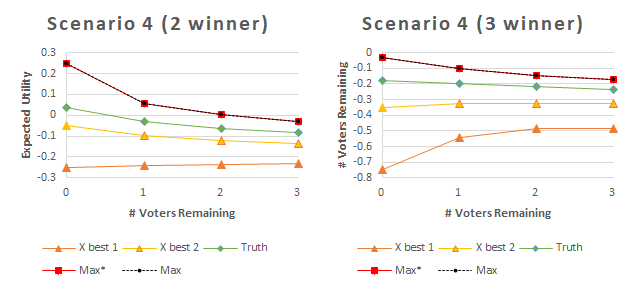}
        \caption{Simulation results for scenario 4}
        \label{fig:sim4}
\end{figure}

Scenario 4 is shown in Figure \ref{fig:sim4}. In this scenario, a negative utility is generated when a disliked candidate wins the election. The simulation results were interesting in that neither being truthful or using the \textit{take the X best} heuristic was the optimal strategy. In both conditions of this scenario, it was preferable to vote for all candidates with a higher utility than the disliked candidate, including those with no utility. Voting in this way resulted in the maximum expected utility and is also the static maximization profile. Voting truthfully also resulted in a higher utility over using the \textit{take the X best} heuristic.

The simulations in Study 2 highlight how important understanding both the voting profile as well as the voting environment is to fully understanding voting behavior in approval elections.  In the scenarios above, using the \textit{take the X best} heuristic was often preferable to other strategies. Using a truthful strategy was preferable only when there was a very small chance to elect any of the candidates in the voters preferred set. In that case, voting for all preferred candidates increases likelihood that at least one preferred candidate will be elected, resulting in the maximum expected utility. In situations like scenario 4 where electing a candidate could result in a negative utility, voting for the top-X candidates was not enough to obtain an optimal outcome. Instead, it is preferable to vote for as many other candidates as possible, thus reducing the likelihood that the disliked candidate will win. Voting for an optimal profile when uncertainty changes (i.e. unknown votes are added afterwards) was outperformed by heuristics in some circumstances (Scenario 1 and 2), but proved to be a superior strategy in others (Scenario 4 and 3 winner Scenario 3).

\section{Conclusions and Future Work}
In this paper we studied the role of heuristics in multi-winner approval voting elections  with and without uncertainty.
Behavioral results from Study 1 showed that the majority of subjects did not find the optimal strategy. The majority also did not vote truthfully. Instead, a significant number of subjects voted using a heuristic to \textit{take the X best}. Follow the leader was utilized less often, though voting for the leader or a leader and the highest utility candidate was common in some scenarios, particularly when the leading candidates surpassed the subjects' preferred candidates by a wide margin.

Simulation results from Study 2 showed that using the \textit{take the X best} heuristic can result in higher utility in many situations with uncertainty. However, in scenarios 3 and 4 we identified situations where \textit{take the X best} was detrimental to the expected utility. For example, the heuristic failed in situations where there was a low chance of electing a preferred candidate. In this situation, the better strategy would be to vote truthfully. \textit{Take the X best} also failed in situations when it was possible to elect a candidate with a negative utility. Instead, it was better to vote for as many candidates as possible that did not have the negative utility, including those for which the voter has no preference.

Going forward, it would be interesting to explore the voting behavior of subjects in situations with uncertainty and identify whether or not voters continue to use a \textit{take the X best} heuristic, even when it is not effective. Another area of interest is to examine whether or not the \textit{take the X best} heuristic works well under other approval voting rules, such as Proportional Approval Voting and Re-weighted Approval Voting, which have been shown to be computationally complex to manipulate \cite{AGGM+15a}. Finally, we would like to explore the use and effectiveness of heuristics in multi-winner voting rules beyond approval voting, such as the single transferable vote. 

More generally, proponents of fast and frugal heuristics have promoted the idea of ecological rationality which means essentially that the rationality of a decision strategy can only be determined when considered in the environment the decision is being made [10]. If for example a simple rule  that ignores some causally relevant information provides the best solution in a specific environment (e.g. simulations 1 and 2) it is considered ecologically rational, despite violating axioms of normative theories (such as expected utility theory). To advance this idea, taxonomies have been created to map different heuristics to different environments, outlining components of the heuristics (search, stopping, and decision rules) and components of the environment (complexity, structure of uncertainty) to show which heuristics may be more or less useful in which environment. We believe that this approach could prove beneficial to our understanding of human behavior in multi-winner voting moving forward.

% Bibliography
\bibliographystyle{ACM-Reference-Format}
%\bibliography{abb,voting,heuristics}

\end{document}